\documentclass[a4paper]{article}

\usepackage{icrc2013}
\usepackage[english]{babel}

\title{An Analog  Trigger System for Atmospheric Cherenkov Telescopes}

\shorttitle{An Analog Trigger for CTA}

\authors{
M. Barcelo$^{1}$, 
J.A. Barrio$^{2}$, 
O. Blanch Bigas$^{1}$, 
J. Boix$^{1}$, 
C. Delgado$^{3}$, 
D. Herranz$^{2}$, 
R. Lopez-Coto$^{1}$, 
G. Martinez$^{3}$, 
L.A. Tejedor$^{2}$, 
for the CTA Consortium
}

\afiliations{
$^1$ High Energy Physics Institute, 08193, Bellaterra, Barcelona, Spain \\
$^2$ Department of Atomic, Molecular and Nuclear Physics, Complutense University of Madrid, 28040, Madrid, Spain \\
$^3$ Energetic, Environmental and Technological Research Center, 28040, Madrid, Spain \\
}

\email{rlopez@ifae.es}

\abstract{
Arrays of Cherenkov telescopes typically use multi-level trigger schemes to keep the rate of random triggers from the night sky background low. At a first stage, individual telescopes produce a trigger signal from the pixel information in the telescope camera. The final event trigger is then formed by combining trigger signals from several telescopes. In this poster, we present a possible scheme for the Cherenkov Telescope Array telescope trigger, which is based on the analog pulse information of the pixels in a telescope camera. Advanced versions of all components of the system have been produced and working prototypes have been tested, showing  a performance that meets the original specifications. Finally, issues related to integrating the trigger system in a telescope camera and in the whole array will be dealt with.
}


\begin{document}
\maketitle

\section{Introduction}
\label{sec:introduction}

The general camera trigger strategy in current Imaging Atmospheric Cherenkov Telescopes (IACTs) looks for an excess of signal located in a relatively small region of the camera within a few nanosecond time window. This approach allows the trigger rate due to Night Sky Background (NSB) accidentals to be reduced, whereas the trigger efficiency for gamma-like events remains high, due to the compactness in time and space of their associated camera images. While this strategy is simple, there is not a unique hardware implementation and different performances can be achieved depending on its details. In particular, current IACTs like HESS~\cite{Trigger_Hess}, VERITAS~\cite{Trigger_Veritas} and MAGIC~\cite{Trigger_Magic,Sumtrigger_Magic} implement alternative solutions, based on different definitions of what an excess is. 

The implementation described in this paper aims to provide an analog trigger scheme to cameras for Medium Size Telescopes (MST)~\cite{NECTAR} and Large Size Telescopes (LST)~\cite{LST} of the Cherenkov Telescope Array (CTA)~\cite{CTA}. Both LST and MST cameras will be segmented in 7-pixel clusters configured in an hexagonal pattern, so that every cluster has 6 direct neighbors, except for the camera borders. Each cluster will be formed by three basic blocks: the Photo-Multiplier Tubes (dubbed PMTs) and their ancillary circuits, a front-end board for readout (including a trigger system), and a backplane for trigger distribution and additional servicing.

Within this arrangement, the proposed trigger system can easily handle trigger rates up to 10 MHz (see below) with a negligible dead time, well below 1\%. It is also flexible enough to implement the trigger decision based either on the Sum or the Majority trigger schemes described below. It relies on the above-mentioned 7-pixels clustering of the camera, defining the trigger regions as different combinations of the signals from each cluster and its neighbors. 

The paper is organized as follows: in section \ref{sec:architecture} an overview of the system is presented, defining the required trigger functionalities and the hardware modules which implement them. Then, several measurements of the performance achieved with the complete trigger system are shown in section~\ref{sec:performance}. Finally, section \ref{sec:conclusion} gathers the main conclusions of the paper and presents the following steps towards a final and integrated trigger system. Further details of the Analog Trigger System for CTA can be found in~\cite{AT-IEEE}

 \begin{figure}[t]
  \centering
  \includegraphics[width=0.45\textwidth]{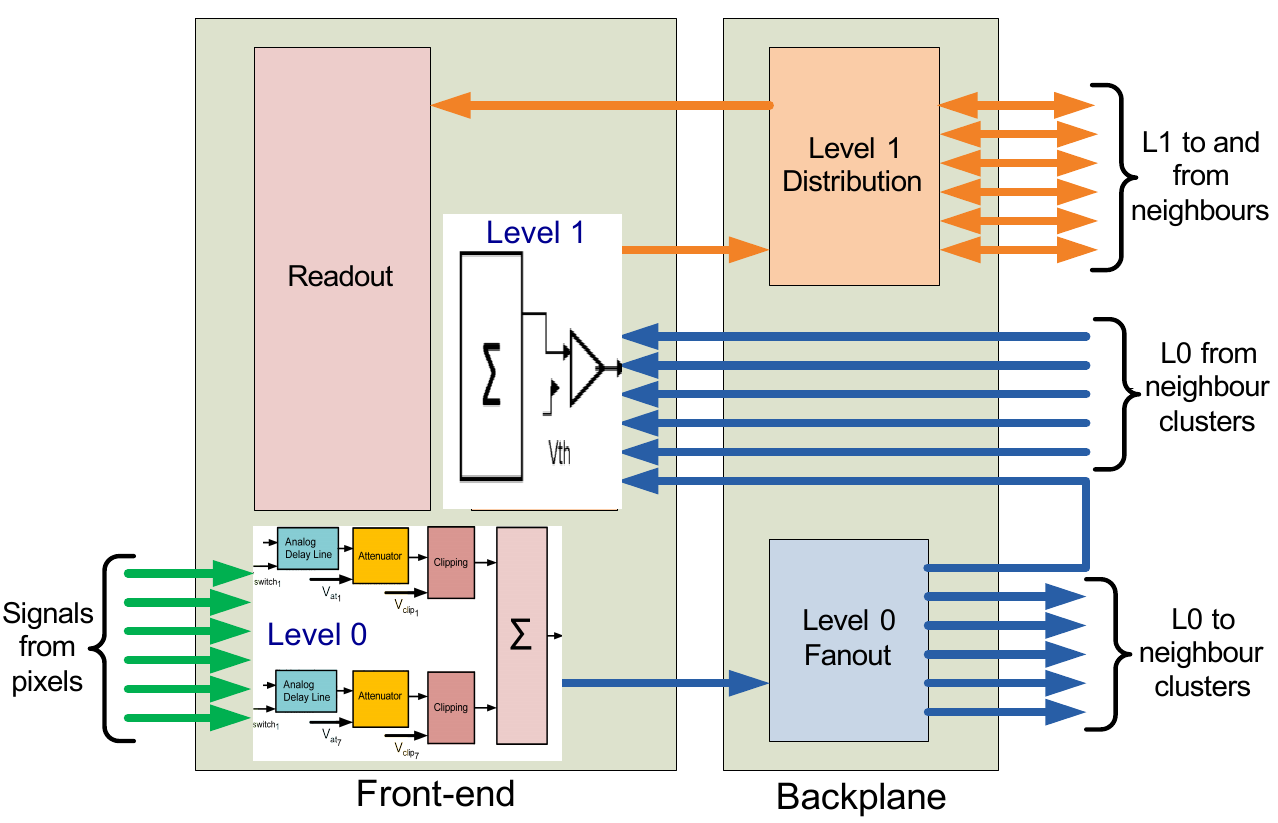}
  \caption{Architecture of the analog trigger system}
  \label{fig:architecture}
 \end{figure}

 \begin{figure*}[!t]
  \centering
  \includegraphics[width=0.77\textwidth]{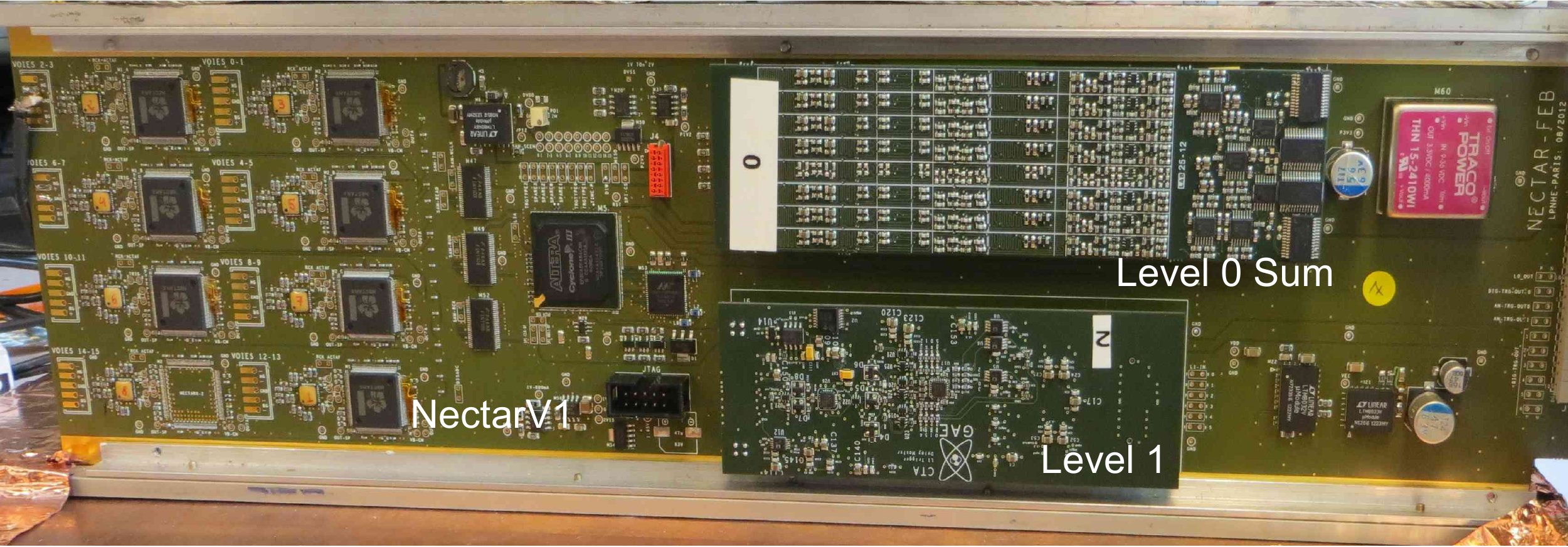}
  \caption{Level 0 and Level 1 mezzanines in a Nectar V1 board}
  \label{fig:l0-l1-nectar}
 \end{figure*}

\section{Architecture}
\label{sec:architecture}

The analog trigger concept is implemented in three stages. In the first one, the so-called Level 0 (hereafter L0) stage combines the analog signals of individual pixels in a cluster into a single analog output signal (whose meaning depends on the type of trigger scheme, see below), which is passed on to the second stage, the so-called Level 1 (hereafter L1). At the L1, the analog addition of the L0 signals from all neighboring clusters  conforming all possible compact regions around a given cluster is examined. The L1 stage takes the decision to trigger the camera if the output of the combination in any cluster of the camera exceeds a given threshold. This decision feeds the third stage, the Level~1 distribution system, which guarantees that the camera trigger signal reaches the readout electronics associated to all clusters at the same time. It also allows to readout arbitrary regions of the camera around the different clusters that gave rise to the L1 signal. With this configuration, the trigger decision (formed by L0 and L1) is as fast as the signal from the Cherenkov light of the air showers, so it will be able to easily handle trigger rates up to 100 MHz. Additionaly, the L1 distribution system introduces a dead time of 40~ns every time the readout is triggered (translating to a trigger rate of ca. 20 MHz), which is negligible to the one introduced by the readout itself.

In its final implementation, the different elements of this trigger system will be placed at different locations on the front-end electronics, as shown in Fig.~\ref{fig:architecture}. While L0 and L1 subsystems are placed in the front-end board itself, the  L1 distribution subsystem it at the backplane of the front-end board, in order to send and receive the trigger signals efficiently among the neighbouring clusters. Current prototypes have been produced as separate mezzanine boards for easy testing, but the final trigger subsystems will lay directly on the front-end and backplane boards. 

\subsection{Level 0}
\label{sec:level0}

The L0 system is responsible for collecting the signals from all pixels in one cluster. These signals are conditioned and then added together before being replicated and sent to the L1 (as described in section \ref{sec:level0-fanout}) of the same cluster and the surrounding ones. In order to test two alternative trigger concepts (Majority and Sum Trigger), two different L0 mezzanines have been developed:

\subsubsection{Majority trigger}

The Majority trigger compares the signal from each pixel with a threshold. If the signal is greater than the threshold, a gate is generated, with its width corresponding to the time-over-threshold. All the gates in a cluster are added analogically, generating a signal with amplitude proportional to the number of pixels above the threshold, which is sent to the L1 subsystem.

Every channel of the Majority trigger mezzanine comprises a differential to single-ended converter and a fast  LVDS (which stands for Low-Voltage Differential Signaling) comparator . The thresholds are generated by a SPI-controlled Digital to Analog Converter (DAC) with 8 outputs. Then, the outputs of the comparators are added in an analog adder, whose output is converted into differential. 

\subsubsection{Sum trigger}

The Sum trigger analogically adds the signals from all pixels in the cluster and sends the resulting signal to the L1 subsystem. Before adding the signals from the individual pixels, each of them goes through attenuator and clipping circuits (both slow-control adjustable). The former allows all pixel gains to be equalized with a precision better than 5\%. The later cuts signals greater than a given value, which limits the influence of after-pulses from the photosensors \cite{Sumtrigger_Magic}. The clipping and attenuator circuits might also reduce the pixel signal to zero, in case that noisy pixels need to be removed from the trigger patterns.

Each channel of the Sum Trigger mezzanine is made up of six different elements. The differential to single-ended converters, the analog adder, the single-ended to differential converters and the DACs are similar to those used for the Majority design, although 2 DACs are required in this case. The attenuator and clipping circuits (exclusive for the Sum modules) are based on PIN diodes and differential Bipolar Junction Transistor amplifiers respectively, and are controlled by DC levels generated in DACs. Fig.~\ref{fig:l0-l1-nectar} shows one of the manufactured L0 sum trigger mezzanine board, connected to a NectarCAM front-end board~\cite{NECTAR}.

Additionally, in the Sum Trigger scheme (and to a lesser extent in the Majority trigger one), the timing of the signals should be very accurate to make the sums (or find the coincidences) properly. In this sense, the different transit times introduced by PMTs with different high voltages can be a problem (as shown in~\cite{Dennis}). To resolve this, the L0 Sum Trigger mezzanine incorporate an analog delay line for each channel, which is formed by a set of switchable passive delay lines, allowing to control the time delay of the analog pulse from 0 to 5.5 ns, in steps of 250 ps.

\subsection{Level 1}
\label{sec:level1}

The L1 system makes several analog sums of the signals from the L0 of the adjacent clusters and compares them with a threshold level. If any of these sums is over the threshold, the L1 sends a trigger signal to the L1 distribution in order to trigger the camera. So, depending on the trigger scheme implemented in the L0, the L1 decides if there are more than a certain number of pixels over the L0 threshold in a region of the camera (Majority Trigger), or if the addition of the photoelectrons generated by all the pixels in this region is greater than the L1 threshold level (Sum Trigger).

The size of the trigger region is an important question to optimize the sensitivity of the telescope for a given configuration. The current design is able to work with trigger regions of 2, 3 or 4 clusters (14, 21 or 28 pixels respectively), which correspond to 3 different operation modes, slow control selectable. There is one L1 subsystem in every cluster, which is in charge of all of the above-mentioned trigger regions. In this way, all possible combinations of 2, 3 or 4 compact clusters are covered in the whole camera, and full trigger overlapping regions are guaranteed. 

The L1 mezzanine is made up of several circuits: first, the differential analog inputs from L0 of the cluster itself and from the neighbouring clusters are scaled and transformed into single-ended signals, to operate with them in a simpler way. Then, the single-ended outputs are split into three or two branches, or not divided at all but attenuated, depending on the specific channel, in order to perform all the required sums for all the trigger regions. Whatever the selected working mode (2, 3 or 4), there are never more than 3 sums required, so all the sums for all the working modes can be performed in a cost-effective way, with only three adders with 4 inputs each. Then, these signals are added and scaled in the adders, which finally send the sums to two different comparators, where the three sums are compared with two different thresholds previously set in DACs. The two comparators (thresholds) are needed in order to allow a partial readout of the camera (as described in~\cite{colibri}). The outputs of the comparators, in LVDS standard, are finally sent to two LVDS OR gates, which provides two differential trigger outputs which will be distributed throughout the camera, activating the readouts of all the clusters. Again in Fig.\ref{fig:l0-l1-nectar}, one of the manufactured L1 mezzanines is shown, connected to a NectarCAM front-end board~\cite{NECTAR}. Finally, a delay calibration system is implemented in the L1 mezzanine, in order to align in time all the analog pulses in a cluster with respect to a reference signal, as described in section~\ref{sec:level0}. Further details of the L1 subsystem can be found in~\cite{Level1-IEEE}

\subsection{Analog Trigger Backplane}
\label{sec:backplane}

\begin{figure}[!h]
  \centering
  \includegraphics[width=0.35\textwidth]{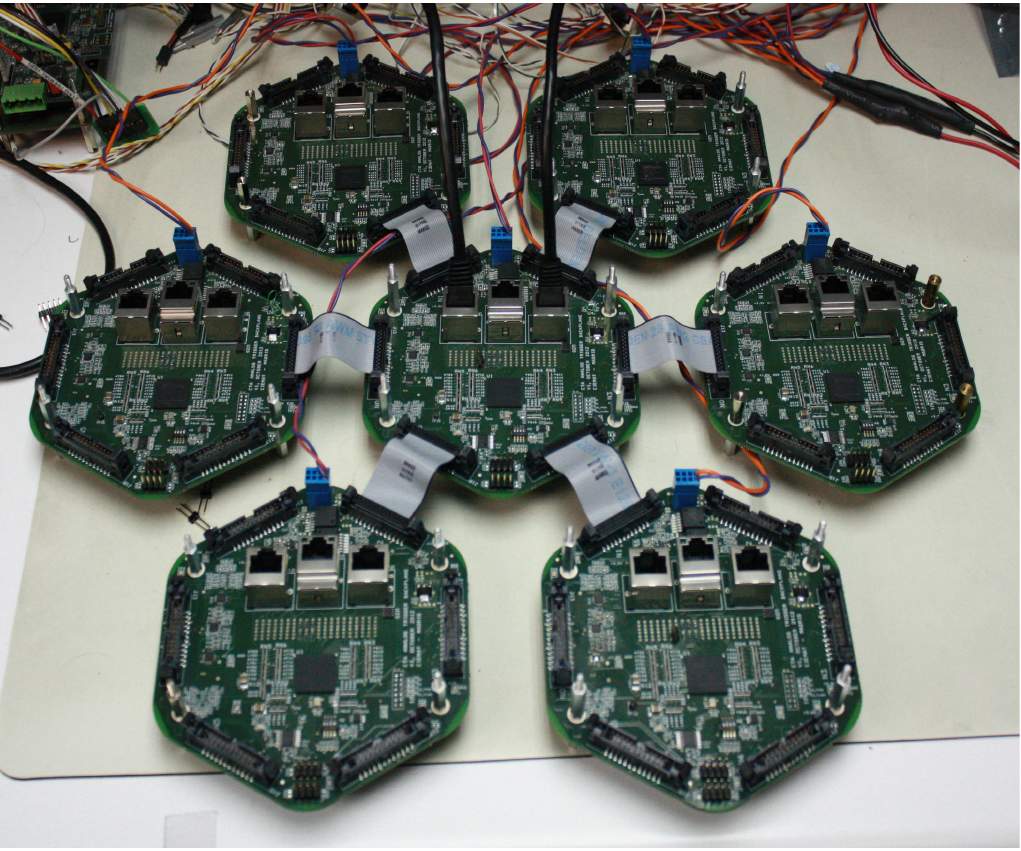}
  \caption{Interconnected Analog Trigger Backplanes}
  \label{fig:backplane}
 \end{figure}

As already mentioned, each cluster includes a backplane board to perform several funtions. A CTA telescope camera which includes the proposed Analog Trigger system must incorporate a backplane board with dedicated trigger functionalities. Therefore, we have designed and prototy\-ped an Analog Trigger Backplane (dubbed ATB) in a single board, which performs, apart from the required trigger functionalities, all the required servicing for the front-end board. Fig.~\ref{fig:backplane} shows 7 such ATB, interconnec\-ted as in a CTA camera.

\subsubsection{Level 1 distribution}
\label{sec:level1-distrib}

The main element of the ATB is the L1 distribution subsystem, whose purpose is the collection of the trigger signals generated by each L1 board, ir order to provide it to all Data Acquisition Systems (DAQ) in all the clusters of the camera. The distributed L1 signal is used to freeze the writing in the DAQ analog memories and start the digitalization process.

In order to save data bandwidth in the camera, the trigger pulse arrival time to all different clusters should be as similar as possible and stable with respect to the L1 generation time during the data taking. Provided this happens, there is no need to read out the whole analog memory but only a subset of cells from each front-end chip. The offset and width of the look-back window can be tuned taking into account the mean and standard deviation of the latency time distribution, that is, the distribution of the elapsed times since the L1 output activation until the L1 trigger pulse is received in the DAQ FPGA.

The proposed distribution system is based on a non-centralized distribution network, which includes a set of distribution modules, one per ATB. Each distribution module is only connected to its 6 immediate neighbors (as shown in Fig.~\ref{fig:backplane}). In addition, each distribution module knows its position in the camera. The main functions of the distribution modules are as follows: reception of the trigger pulse from the L1  module of the corresponding cluster, delaying of the L1 pulse before transmitting it to the neighbors (the length of the delay depends on the distance to the central cluster), transmission and reception of the L1 pulse to the appropriate neighbors depending on the cluster position in the camera and finally transmission of the L1 pulse (after its corresponding delay, which depends on the cluster position) to the DAQ FPGA.

The reception and transmission of the L1 pulses during the distribution has to be asynchronous, in the sense that the L1 signal is not synchronized with any clock during its processing and routing along the distribution path. Thus, the event time information encoded in the leading edge of the L1 decision output is preserved.

The minimum value of latency of the distribution system is defined by the worst case time needed to carry out the transmission of the L1 signal from one cluster to the others. The distribution system has to be calibrated in order to equalize the distribution times to be equal to the worst case.

The proposed implementation for this distribution scheme is based on low cost FPGAs. These devices are compatible with the interfaces of the system and provide a versatile and reconfigurable way for the implementation of the basic functions that have to be performed: LVDS reception and transmission, routing logic and delay lines. One low cost FPGA Xilinx Spartan 6 is used for each distribution module. 

This distribution scheme has been successfully tested with mezzanine boards plugged in a general purpose backplane (see~\cite{AT-IEEE} for further details) and has recently been integrated in the ATBs shown in Fig.~\ref{fig:backplane}.

\subsubsection{Level 0 fan-out}
\label{sec:level0-fanout}

The L0 output signal of a given cluster must be distributed to the neighbouring clusters in order to perform all possible additions of clusters in the L1 subsystem. To do so, the L0 output signal of each cluster is sent to the analog branch of the ATB, where it is fanned-out to several branches corresponding to the neighbouring backplanes of the clusters which have to receive that signal, and one more copy for the cluster itself. In the same way, every cluster backplane receives L0 signals from its neighbours, which are then sent to the L1 subsystem, along with the L0 signal from the cluster itself. 

In order to achieve this functionality, the fan-out system is implemented using differential signals, in order to minimize the effect of noise in the long transmission lines. Additionally, lumped element Wilkinson dividers and amplifiers have been used to obtain similar gains and good matching features for the all the six outputs. Finally, a switch is implemented in the ATB which either sends the L0 signal to the fan-out system or redirects it back to the L1 subsystem. This functionality is required to perform the above-mentioned calibration of the analog delay lines. In the same was as for the L1 distribution, the L0 fan-out scheme has been successfully tested with mezzanine boards plugged in a general purpose backplane (see~\cite{AT-IEEE} for details) and has recently been integrated in the ATBs shown in Fig.~\ref{fig:backplane}.

\subsubsection{Other functionalities}

The ATB is the only backplane board of the cluster, so it must deliver all the additional servicing required for normal operation of the cluster, such as providing the front-end board with Ethernet (for data transfer) and power interfaces, as well as IP address. It also includes a way to identify the physical location of the clusters in the camera. Finally, it uses the L1 distribution system to deliver the clock signals required by the front-end digitizers.

\section{Performance}
\label{sec:performance}

A number of prototypes of each element in the Analog Trigger System have been produced and individually characterized in stand-alone test benches. Additionally, several tests have been performed by integrating all the subsys\-tems, showing that the performance of the system meets the predefined specifications. Finally, integrated measurements with both NECTAR~\cite{NECTAR} and DRAGON~\cite{DRAGON} readout boards have been done, demonstrating good mechanical, electrical and logical compatibility.

\begin{figure}[!t]
  \centering
  \includegraphics[width=0.35\textwidth]{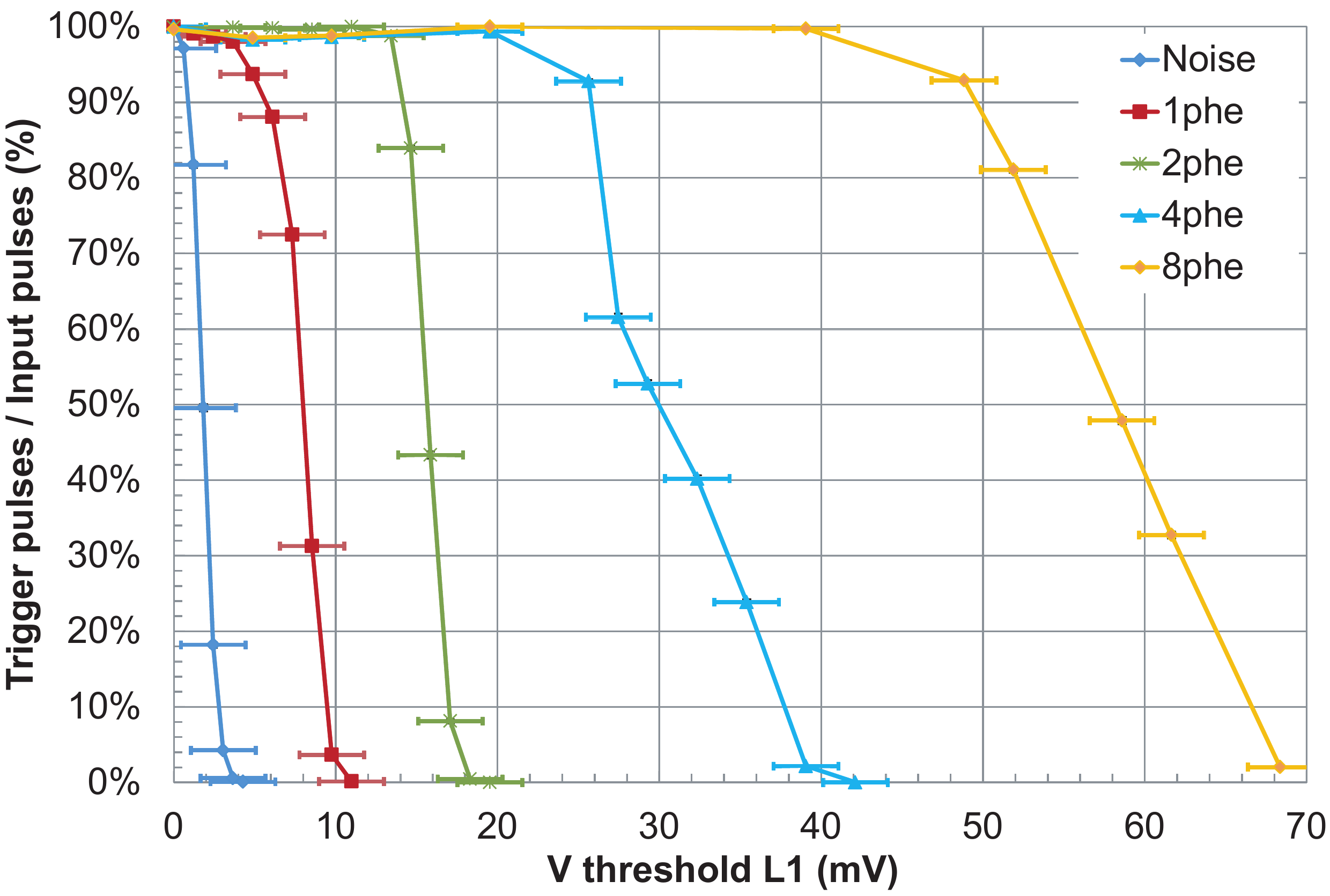}
  \caption{Example of performance.}
  \label{fig:performance}
 \end{figure}

The Analog Trigger has a maximum differential output voltage of 2 V for a maximum number of 200 photo-electrons (dubbed phe). The electronic noise has been also measured, proving to be below 2 mV per cluster taking part in the trigger decision. Therefore the electronic noise is at the level of 0.2 phe (as shown in fig.~\ref{fig:performance}), well below the NSB, which even for a dark patch will be above 2 phe per cluster. Moreover, the achieved dynamic range allows the threshold to be calibrated with calibration optical pulses such as the ones in use for Cherenkov telescopes.

The analogue memories must have a register deep enough to avoid over-writing during the time that the trigger system takes to decide and distribute the trigger signal (i.e. the trigger latency). In the case of the LSTs, the most significant part of the trigger latency comes from the exchange of L1 triggers among telescopes, while for MSTs it is the camera trigger latency itself that dominates. So, the latency of the whole Analog Trigger System has been measured to a total value of 190 ns, the L0 and L1 contributing with 20 ns and the L1 distribution contributing with 170 ns.

The construction of the prototypes have also allowed to obtain a precise estimation of the power consumption and weight for the Analog Trigger System. The power consumed by the L0 and L1 is below 590 mW per channel, although nearly one third corresponds to the clipping circuits, so if the clipping stage can be finally avoided by choosing  PMTs with very low after-pulsing, the final power consumption could go down to 400 mW. The power consumed by the ATB is around 200 mW per channel. The weight of the total system is around 25 g per channel, 4 g corresponding to the L0 and L1 (taking into account only their components) and 21 g to the ATB.

\section{Conclusion and outlook}
\label{sec:conclusion}

A complete Analog Trigger System for the CTA individual telescopes has been presented, which is compatible with the modular design of the telescopes, therefore simplifying the hardware and cabling. The analog treatment of the signals allows to implement a sum trigger scheme in addition to the classic majority, improving the sensitivity to gamma-rays with energies below 200 GeV. 

The different modules that make up the trigger system have been manufactured as mezzanines and tested together with the other electronic components of the clusters, meeting the expected requirements. Therefore, they are ready for integration and mass production if they were chosen as the trigger system for the LSTs and MSTs of CTA.

Finally, there is still room for further improvement in Analog Trigger system, namely its implementation with the full functionalities described above, in the form of two ASIC circuits, one for the L0 (including both Sum and Majority trigger schemes) and another one for the L1. In this way, a drastic reduction in cost, weight, space and power consumption can be achieved. This implementation is currently in its design phase, with the first prototypes to be expected by the summer 2013.

\vspace*{0.5cm}
\footnotesize{{\bf Acknowledgment:}{
We would like to thank our colleagues from NECTAr and DRAGON teams for their collaboration in the characterization of the trigger system, and M. Punch for fruitful discussions. We gratefully acknowledge support from the agencies and organizations listed in this page: \url{http://www.cta-observatory.org/?q=node/22}
}

\end{document}